\documentclass[aps,twocolumn,superscriptaddress]{revtex4}

\usepackage[dvips]{graphicx}
\usepackage{dcolumn}
\usepackage{bm}
\usepackage{amsmath}
\usepackage{latexsym}

\begin{document}


\title{Electron beam driven alkali metal atom source for loading a magneto-optical trap in a cryogenic environment}

\author{S. Haslinger}
 \email{haslinger@ati.ac.at}
 \homepage[URL: ]{http:\\www.ati.ac.at}
 \affiliation{Vienna Center for Quantum Science and Technology, Atominstitut, TU-Wien, 1020 Vienna, Austria}
\author{R. Ams\"uss}
\author{Ch. Koller}
\author{C. Hufnagel}
 \affiliation{Vienna Center for Quantum Science and Technology, Atominstitut, TU-Wien, 1020
Vienna, Austria}
\author{N. Lippok}
 \affiliation{Vienna Center for Quantum Science and Technology, Atominstitut, TU-Wien, 1020 Vienna, Austria}
 \affiliation{Planet and Star Formation Department, Max-Planck-Institut for Astronomy, 69117 Heidelberg, Germany}
\author{J. Majer}
 \affiliation{Vienna Center for Quantum Science and Technology, Atominstitut, TU-Wien, 1020 Vienna, Austria}
 \author{J. Verdu}
 \affiliation{Vienna Center for Quantum Science and Technology, Atominstitut, TU-Wien, 1020 Vienna, Austria}
 \affiliation{Department of Physics and Astronomy, University of Sussex, Brighton, UK}
 \author{S. Schneider}
 \author{J. Schmiedmayer}
 \affiliation{Vienna Center for Quantum Science and Technology, Atominstitut, TU-Wien, 1020 Vienna, Austria}

\date{\today}

\begin{abstract}
We present a versatile and compact electron beam driven source for alkali metal atoms, which can be implemented in cryostats. With a heat load of less than 10mW, the heat dissipation normalized to the atoms loaded into the magneto-optical Trap (MOT), is about a factor 1000 smaller than for a typical alkali metal dispenser.  The measured linear scaling of the MOT loading rate with electron current observed in the experiments, indicates that electron stimulated desorption is the corresponding mechanism to release the atoms.
\end{abstract}


\maketitle

Preparing ensembles of ultra-cold atoms with a magneto-optical Trap (MOT) has become a standard technique in atomic physics and is the first major step on the way to Bose-Einstein condensation (BEC) \cite{Nob1,Nob2} and ultra-cold quantum gases. Typical sources for alkali metal atoms used in BEC-experiments \cite{And95,Dav95,Gri03,Ing01,Hul97} are among others, alkali metal ovens \cite{Hadzi2002} feeding Zeeman slowers \cite{Bar91} or alkali metal dispensers \cite{SAES} generating vapour of the alkali atoms from where the MOT is loaded. The latter use resistive heating to chemically reduce compounds of alkalies to produce the atomic vapor.

Recently, the increasing interest for studying the interaction between ultra-cold quantum gases and solid state quantum devices \cite{Schoelkopf:2008k,Ver09,SvdWCL:04,Petrosyan:08,Petrosyan:09}, brought up a need for alkali metal sources compatible with the thermal load in cryogenic environments where superconducting quantum devices operate. As the ideal platform for such experiments is an atom chip \cite{Folman2000,Folman2002,Reichel2002,Fortagh2007}, state of the art experiments with ultra-cold atoms on superconducting atom chips, use sophisticated transport schemes \cite{Nirrengarten2006,mukai07,Roux08,Kas09,Huf09,Emm09}. Implementations are realized by transferring pre-cooled trapped atoms into the cryostat using a moveable magnetic trap \cite{mukai07}, optical tweezers \cite{Cano1} or push a slow beam of atoms into a cryogenic system with an integrated MOT \cite{Roux08}.

In the following, we present a versatile and compact electron beam driven cold atom source which is compatible with a cryogenic environment. Depending on the preparation of the field emission tip, it is possible to set free a significant amount of trappable atoms at a few 100$\mu$W of heat load. We observe MOT loading rates which, when scaled by total heat load to the system, are a factor of 1000 higher than for a typical MOT loaded by an alkali metal dispenser. In a typical experiment, we load $3\times 10^6$ atoms within 1.5s at a total heat load of 8.4mW. Compared with other atom sources conceived for cryogenic systems such as light induced atomic desorption (LIAD) \cite{Moi94,Tom03}, or laser ablation \cite{Doy95}, this atom source is fully tunable as it relies on an electron beam, rather than on laser light as input power source.

\begin{figure}[h]
\includegraphics[width=\columnwidth]{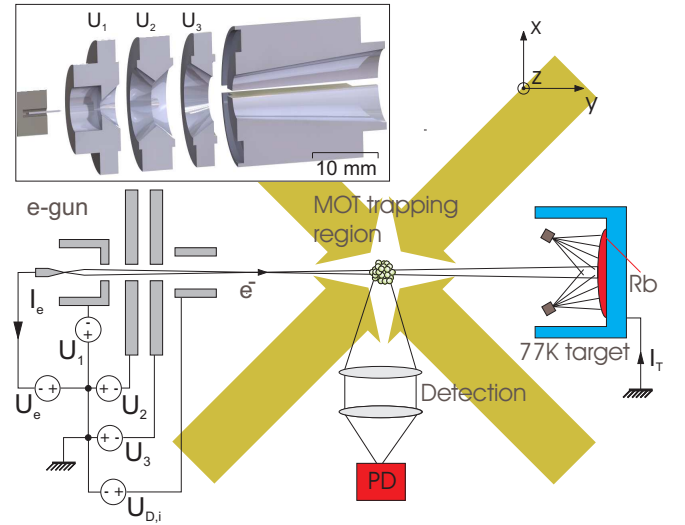}
\caption{Schematic of the cryogenic e-beam driven alkali atom source for loading a MOT: The electron beam emerges from a cold field-emitter, crossing the trapping region with the crossed laser beams and hitting a liquid nitrogen cooled rubidium target. Two alkali metal dispensers are used to prepare an alkali metal layer on the surface. Released $^{87}$Rb atoms are loaded into the MOT and the fluorescence of the trapped atoms is measured with a photodiode (PD). The inset shows a 3D image of the conical deflection plates and the high efficient cold field-emitter source. The kinetic energy of the electrons is given by the emission voltage $U_e$.}
\label{fig_1}
\end{figure}

The setup of an electron beam driven atom source is shown in Fig.\ref{fig_1}. An electron beam is created by a field emission source. The electron beam is then directed onto a cryogenic Rb target at 77K. Using PtIr or tungsten field emission tips, we achieve a high flux of laser trappable rubidium atoms at maximum target heat loads of $<$10mW. The Rb atoms desorbed from the target by the electron impact are trapped in a close by magneto optical Trap, situated in front of the target at a distance of 7cm.

The electron beam is prepared by a field-emission tip which avoids thermal radiation from the electron source. We used both, PtIr or tungsten tips. With either of the two tips, the source is capable of producing a beam current of more than 10$\mu$A at kinetic energies up to 6keV. To preserve the tip, the field-emitter is operated at currents below 10$\mu$A. 

With a system of electrostatic lenses, the electron beam is focused onto the target to a spot size of $\leq$600$\mu$m diameter. The design of the lens system is based on \cite{Suz1,Suz2}, where particular attention was paid to optimize efficient transmission via minimizing the loss of the emitted current $I_e$ propagating through the lens system. The transfer efficiency $I_T$/$I_e$ of the target current $I_T$ and the emitted current is up to 0.4. The other electrons hit the electrodes of the lens system and do not reach the target \footnote{As simulation software for the optimization of the electron gun and the lens system, the COMSOL Multiphysics Modeling and Simulation package was used.}.
We characterized the electron beam in previous experiments, where the target was replaced with a phosphor screen. Measuring the beam profile, beam position and beam current, we used the deflection plates to minimize the effect of the magnetic quadrupole field of the MOT on the electron beam. In addition, the four conical deflection plates allow the beam to be adjusted in x- and z-direction up to 130 mrad (see Fig.\ref{fig_1}), applying a voltage of $U_{D,i=1:4}$ at each of the plates.
The target current $I_T$ and the emission current $I_e$ are continuously monitored.

Leaving the field-emission source, the electrons cross the MOT region and hit the liquid nitrogen cooled target. Thereby the probability of ionizing the Rb \cite{Ger08,Sch96} in the MOT is much less than $1\%$ for the kinetic energies of the electrons, the short loading times and the low atom density, and can therefore be neglected.\\
Initially our Rb target is prepared the following way: Rb is deposited on the oxidized Cu surface from two very close by (see Fig.\ref{fig_1}) alkali metal dispensers. The dispensers \cite{SAES} are switched on for 60s with $2 \times 5A$, and coat the $LN_2$ cooled target surface with a thick Rb film as most of the Rb sticks on the Cu surface (partial pressure of Rb at 77K is  about $10^{-15}$mbar \cite{Ste01}). The target lasts for more than 1000 experimental cycles.

To characterize the vacuum during target preparation and the experimental cycle, we observe the overall pressure in the chamber with an UHV gauge and the partial pressure of $^{87}\text{Rb}$ and other rest gas species using a mass spectrometer. During a typical experimental cycle the total pressure is $ < 2\times 10^{-10}$ mbar.

When the electrons hit the target, neutral rubidium atoms are desorbed and the low velocity tail of the released $^{87}\text{Rb}$ atoms can be trapped and laser cooled in the MOT. We operate the MOT using a diode laser, $\delta=-18$MHz detuned from the $5^2S_{1/2}(\text{F}=2)\rightarrow5^2P_{3/2}(\text{F}'=3)$ transition at 780 nm and a typical quadrupole gradient of 20G/cm. A second diode laser tuned to the $5^2S_{1/2}(\text{F}=1)\rightarrow5^2P_{3/2}(\text{F}'=2)$ transition pumps atoms back from the $\text{F}=\text {1}$  to the $\text{F}=2$ hyperfine state. In each of the three retro-reflected trapping beams we employ $\approx$20mW of laser power. At the end of each experimental phase the MOT laser frequency is ramped through resonance ($\delta=0$) to 5MHz blue detuning within 2ms. The number of collected atoms can be calculated via fluorescence from the current peak value of the photodiode at resonance $\delta=0$, where the signal of the calibrated photodiode is proportional to the number of trapped atoms \cite{Raa87}. Ramping the detuning into the blue, expells all atoms from the trap and resets the number of trapped atoms to zero. 

\begin{figure}[t]
 \includegraphics[width=\columnwidth]{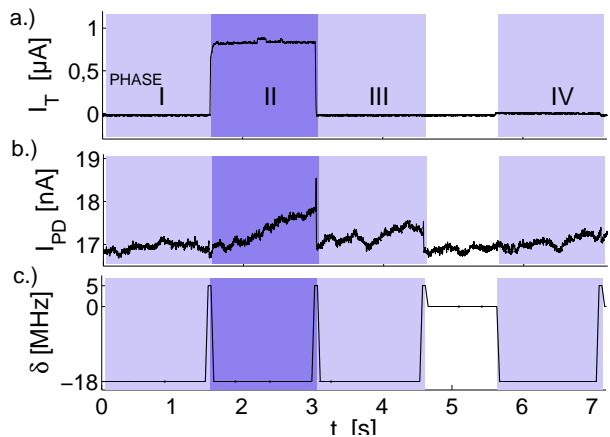}
 \caption{a.) Electron beam current at target. 
 b.) Fluorescence signal $I_{PD}$ from the photo diode, from which the number of trapped atoms is determined. The signal shows resonance peaks after 1.5, 3, 4.6 and 7.1 seconds, when the cooling laser is ramped over resonance.  
 c.) At the end of every MOT phase, lasting 1.5s, the 780nm cooling laser at the $5^2S_{1/2}(\text{F}=2)\rightarrow5^2P_{3/2}(\text{F}'=3)$ transition is ramped from $\delta=-18$MHz to $\delta=+5$MHz within 10ms. During the pause between phase III and IV the laser stays at resonance $\delta=0$.
\label{fig_2}}
\end{figure}

Fig.\ref{fig_2} shows the experimental cycle which consists of phase I-IV. We characterize our Rb source by means of a measurement cycle where the MOT is loaded in each phase, lasting 1500ms.\\
In phase I, the MOT is loaded without the electron beam switched on, in order to get a reference measurement of the background.\\
In phase II the electron beam is switched on and the desorbed atoms are trapped in the MOT. At the end of phase II the electron beam is switched off by setting the emission voltage to zero.\\
During phase III, we measure how many atoms are loaded from the remaining background gas while the e-gun is off, where $N_{\text{III}}$ is in the order of $<$0.4$\times$ $N_{\text{II}}$.\\
In phase IV, we switch the electron emitter on, but block the electron beam such that it can not reach the target. This is achieved by applying a large negative voltage $U_3$ at the blocking electrode, which is ramped up between phase III and phase IV. Technical limitation for the ramping speed of the blocking voltage lead to a pause between phase III and phase IV. In phase IV the beam is not hitting the target, eventhough a small leakage current remains. This phase therefore allows us to assess that the trapped $^{87}\text{Rb}$ in the MOT origins from the target rather than being released from other surfaces accidentally hit by electrons.\\    
After phase IV, the cycle is paused for 13 seconds before restarting with phase I. The total experimental cycle lasts 20 seconds. 

During the operation of the electron gun the first electrode is biased with $U_1=-\text{500V}$ and the emission voltage $U_e$ is varied from -2.5kV up to -3.0kV to adjust the electron beam current which is related to the emission voltage $U_e$ by the Nordheim-Fowler equation \cite{NoF28}. The focusing voltage $U_2$ is set to $-440$V.

\begin{figure}[t]
 \includegraphics[width=\columnwidth]{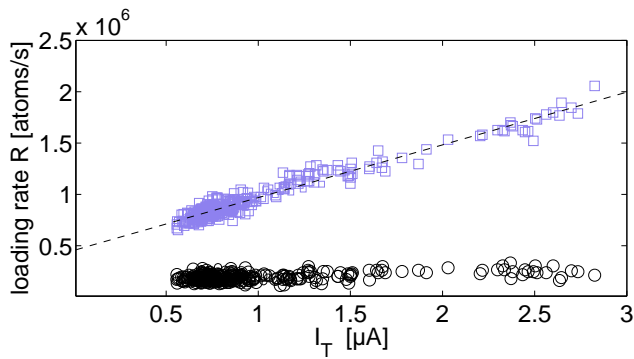}
 \caption{
A typical measurement of the loading rate ($\Box$) in dependence of the target current $I_T$. The atoms are desorbed during phase II, and loaded with a yield $Y_D$ proportional to the slope of the dashed line. A reference measurement ($\circ$), from phase I shows the rate of loading the MOT from the background.
\label{fig_3}}
\end{figure}

Operated with a PtIr emission tip, we observe that an increase of the electron current on the target $I_T$ results in a linear increase in the number of trapped atoms $N_{\text{II}}$ during phase II and hence in the loading rate R (see Fig.\ref{fig_3}). In addition we measured no dependence of the trapped atoms on the electron beam spot size. 
In a typical experiment, we trap about $3\times 10^6$ atoms at a target current of $2.8\mu$A after a loading time $t_L$ of 1.5s and $4.5\times 10^6$ atoms after $t_L=2.5$s. In a different measurement were we loaded the MOT for up to 20sec, we find a characteristic time constant of about 10s, fitting an exponential law to the loading curve of the MOT. This loading time constant is much longer than the loading time $t_L=1.5$s during phase I-IV. This results in a near linear behavior of the loading curve, and the loading rate can be estimated from the number of trapped atoms divided by $t_L$. We determine a loading yield of $Y_D=5.1\times 10^5$ atoms/s/$\mu$A, which gives the number of atoms in the MOT per $\mu$A electron target current $I_T$ and which is proportional to the slope of the curve in Fig. \ref{fig_3}.\\
For a tungsten emitter operated at 650V and using a divergent electron beam with cross section of several $cm^2$ on the target, we found a loading rate of $ R=7.3\times 10^5$ atoms/s at a total target power of only $200\mu$W. A conventional MOT operated in the same chamber under identical laser and vacuum conditions (background pressure of $1.2\times 10^{-10}$mbar), loaded from a Rb-dispenser, yields a loading rate of $\approx 1.9\times 10^7$ atoms/s using a resistive power of $\approx$18W \footnote{In addition, we point out that Rb alkali metal dispensers demand a minimum power in the order of at least 8W to set free a feasible amount of Rb.}. Comparing both, we see that the electron beam Rb source achieves a good yield of trapped atoms even at very low powers. The total heat load from the e-beam driven Rb source, scaled by the MOT loading rate, is more than a factor 1000 lower than for loading from a Rb dispenser.\\

To explain the experimental observations, we consider two potential mechanisms to desorb neutral alkali metal atoms from an oxidized metal surface as a result of electron impact: (1) electron stimulated desorption (ESD) and (2) thermal desorption due to electron beam heating. While the first effect relies on the neutralization of adsorbed alkali metal ions by a charge transfer process from the oxidized surface, in the latter case a part of the electrons kinetic energy is converted into heat and leads to thermal evaporation of surface atoms.

According to Ageev et al. who studied ESD of Li, Na, K and Cs from alkali metal layers on oxidized tungsten \cite{Age95}, an incident electron creates a core-hole in the oxygen 2s level which stimulates an intra-atomic Auger-decay and allows for a subsequent neutralization of a positive alkali metal ion. If the positive oxygen ion can capture electrons from the substrate to achieve a negative charge state again, the alkali metal atom will be repelled and desorb as a neutral atom. This desorption process will increase the partial pressure of an adsorbed species \textit{j}, and can be described by  
\begin{equation}
	\Delta \dot{Q}_j=(p^{1}_j-p^{0}_j)S_j=\eta_j \frac{I_e}{e}k_B T
	\label{1}
\end{equation} 
as written in \cite{Tratnik2005}. 
Here, $\Delta \dot{Q}_j$ is the differential rate for desorbing particles of species $j$, $p^{0}_j$ and $p^{1}_j$ are the steady-state partial pressures before and after electron impact, $S_j$ is the effective pumping speed of species $j$, $\eta_j$ the molecular desorption yield, $I_e$ the total electron current, $k_B$ the Boltzmann constant and $T$ the temperature of the target. Equation (\ref{1}) assumes that the electron current density is low compared to the adatom density. This leads to the observed linear relationship between the desorption rate and the electron current (Fig.~\ref{fig_3}). 

We also established a simple mathematical model for the thermal desorption of a thin layer of Rb due to electron impact. Following a model from Lin et al. \cite{Lin67}, the local temperature rise due to the electron beam on the Rb layer can be estimated. With the temperature known, the Langmuir-Knudsen law \cite{Lang1913} describes the presented desorbing process and mass flow. Assuming a thermal velocity distribution and a maximum capture velocity of the MOT, we obtain estimates for the number of trapped Rb atoms in the MOT. Due to the rapidly increasing vapor pressure of Rb with increasing target temperature, one expects a highly non-linear, exponentially shaped relationship between the number of trapped atoms and the electron current on the target in contradiction to our observation (Fig.~\ref{fig_3}). 

We deduce from these simple models that the linear relationship between the number of desorbed atoms and the target current is rather described by ESD than a thermal desorption process. In order to substantiate this, further studies e.g. at lower electron energies would be necessary, which wasn't possible with our electron source. In addition, we would also like to point out that we observe an increase in the partial pressures of rest gas species ($\text{H}_2$, $\text{CO}_2$, and $\text{N}_2$) in our chamber, during operation of the electron source. This can be understood by ESD of non-metals, induced by stray electrons hitting a surface different from our target. Due to the increased background pressure we were constrained to short MOT loading times. 

In conclusion, we present an electron beam driven source for Rb atoms desorbed from a 77K target, which is able to load a magneto-optical trap. The electron beam is emitted either from a PtIr-tip or an etched W-tip where the PtIr-tip is easier to produce and the W-tip demands less emission voltage to be operated. The linear dependence of the MOT loading rate with the electron current impinging on the target depends weakly on the focused spot size of the beam. In addition, the low power density of the divergent beam suggests that electron stimulated desorption and not thermal evaporation is the mechanism to release the atoms from the surface. 
With its low power needed to operate, the atom source does not present a significant heat load for a cryogenic cold atom experiment. Our electron beam driven Rb source requires more than a factor 1000 less input power to load a MOT when compared to standard Rb dispensers under the same conditions. This demonstrates that electron beam driven atom sources can provide several $10^6$ trappable atoms in cryogenic environments with low cooling powers.

We thank T. Juffmann, Universit\"at Wien for supporting us with tungsten emitters, and J. Summhammer and M. Fugger for technical support.
This work was supported by the European Union project MIDAS and the Austrian Science Fund FWF. SH acknowledges support from the DOC program of the Austrian Academy of Science (\"OAW), CK from the FUNMAT research alliance, RA and NL from the COQUS doctoral program. JM acknowledges support from the Marie Curie Action HQS.\\
S. Haslinger and R. Ams\"uss both contributed equally to this work.


\bibliographystyle{apsrev}

\end{document}